\documentclass{article}

\usepackage[utf8]{inputenc}
\usepackage{amsmath}
\usepackage{verbatim}
\usepackage{amsthm}
\usepackage{amssymb}
\usepackage{float}
\usepackage{algorithm}
\usepackage{algcompatible}
\usepackage[noend]{algpseudocode}
\usepackage{graphicx}
\usepackage{hyperref} 
\usepackage{enumerate}
\usepackage{multicol}
\usepackage{array}
\usepackage{tikz}
\usepackage{longtable}
\usepackage[bottom=1.2cm,top=1cm,left=1.0cm,right=1.0cm]{geometry}
\graphicspath{{img2/}}
\usepackage[outdir=./img2/]{epstopdf}

\newcommand{\G}{$G=(V,E)$ }
\newcommand{\raiz}{\text{$v_{0}$} }
\newcommand{\Gv}{$H=(V,E,\raiz,L,\Omega)$ }

\newcommand{\nA}{\textbf{\textit{A}}}
\newcommand{\nC}{\textbf{\textit{C}}}
\newcommand{\nJ}{\textbf{\textit{J}}}

\newcommand{\VB}{\text{$v_{B}$}}
\newcommand{\VA}{\text{$v_{A}$}}
\newcommand{\VL}{\text{$v_{L}$}}
\newcommand{\VH}{\text{$v_{H}$}}

\algblock[Try]{Try}{EndTry}
\algnotext[Try]{EndTry}
\algrenewtext{Try}{\textbf{try}}

\algblockdefx[Catch]{Catch}{EndCatch}%
[1][]{\textbf{catch} #1}

\algnotext[Catch]{EndCatch}
\newcommand{\inicioAlgoritmo}[1]{

\begin{algorithm}[H]
\caption{#1}
\begin{algorithmic}[1]}

\newcommand{\fimAlgoritmo[1]}{\end{algorithmic}
\end{algorithm}}

\algnewcommand\algorithmicinput{\textbf{Input:}}
\algnewcommand\Input{\item[\algorithmicinput]}
\algnewcommand\algorithmicoutput{\textbf{Output:}}
\algnewcommand\Output{\item[\algorithmicoutput]}

\newtheorem{definition}{Definition}
\newtheorem{theorem}{Theorem}

\newtheorem{prop}{Property}

\newtheorem{conjecture}{Conjecture}

\begin{document}

\title{Emergent chaotic iterations in hard sparse instances of hamiltonian path problem}
\author{Cícero A. de Lima \footnote{email: cicero.lima.id.1@gmail.com, orcid: 0000-0002-3117-3065}}
\maketitle

\begin{abstract}

A hamiltonian path is a path walk $P$ that can be a hamiltonian path or hamiltonian circuit. Determining whether such hamiltonian path exists in a given graph \G is a NP-Complete problem. In this paper, a novel  algorithm with chaotic behaviour for hamiltonian path problem is proposed. We show that our algorithm runs in $O(V^5(V+E))$ for hard sparse instances from FHCP challenge dataset.
\end{abstract}

\section{Introduction}

\paragraph{Contributions} In this paper, we provide a implementation of SFCM-R algorithm, which is, by our knowledge, the first algorithm that relies on emergent chaotic iterations to simulate a chaotic turing machine capable of finding hamiltonian path in a variety of hard instances of HCP in polynomial time/space, thus contributing for a better understanding of the hamiltonian path problem and NP-hard problems in general in the light of dynamical systems.

\paragraph{Diagnose Tools} In this paper, we use 0-1 test and Lyapunov expoent $\lambda$ and 0-1 test to test whether SFCM-R output is a dynamical chaotic system. Lyapunov expoent indicates chaos when $\lambda>0$ where nearby trajectories separate exponentially. The 0-1 test provides a single statistic, which approaches 1 for chaotic systems. Both take as input a time-series data of measurements $0 \leq x \leq 1$ corresponding to difference degree between algorithm state and solution, where 0 is full similarity to solution.

\section{Preliminary}
\paragraph{Conventions} By convention, $v_{LABEL}$ is a vertex labelled as $v_{LABEL}$ and $N_{v_{LABEL}}(w)$ represents a set of vertices $w^\prime \in N(w)$ labelled as $v_{LABEL}$. $H^\prime_\raiz$ is the root of $H^\prime \supseteq H$ and $H^\prime_v$ is the $v$ of its current state. The notation $H_{v_{LABEL}}$ represents a set of all vertices labelled as $v_{LABEL}$. Before continuing, we'll briefly describe what each label means. Let $w$ be a vertex $w \in V$. If $w$ is an articulation point of $\tau \in \Omega$, it will labelled as $\VA$ or \textit{minimal articulation} vertex. A vertex labelled as $\VA$ or $\VL$ is a $\VB$ vertex. For conciseness, Every real articulation point of $H$ is labelled as \VH\space and $C_\VH=\VH \cup N(\VH)$ represents a $C_\VH$ component. 

\paragraph{Mathematical background} Let us denote by $[[a;b]]$ the following integers: ${a,a+1, ... , b}$, where $a<b$. A \textit{system} under consideration iteratively modifies a collection of $n$ components. Each component $i \in [[1;n]]$ takes its value $x_i$ among the domain $\mathbb{B} = \{0,1\}$. A \textit{configuration} of the system at discrete time \textit{t} is the vector $x^t = (x^{t}_{1},..., x^{t}_{n} ) \in \mathbb{B}^n$. The dynamics of the system is described according to a function $f = \mathbb{B}^n \to \mathbb{B}^n$ such that $f(x)= f_1(x),...,f_n(x))$. Let be given a configuration $x$. In what follows $N(i,x) = (x_1,...,\overline{x}_i,..., x_n)$ is the configuration obtained by switching the $i$th component of $x$ ($\overline{x}_i$ is indeed the negation of $x_i$). Intuitively, $x$ and $N(i,x)$ are neighbours. The discrete iterations of $f$ are represented by the oriented \textit{graph of iterations} $\Gamma(f)$. 

In such graph, vertices are configurations of $\mathbb{B}^n$ and there is an arc labelled $i$ from $x$ to $N(i,x)$ is and only if $f_{i}(x)$ is $N(i,x)$.
In the sequel, the \textit{strategy} $S=(S^t)^{t \in \mathbb{N}}$ is the sequence defining which component to update at time $t$ and $S^t$ denotes $i$th term. This iteration scheme that only modifies one element at each iteration is usually refereed to as \textit{asynchronous iterations}. More precisely, we have for any $i$,$1 \leq 1 \leq n$. 

\begin{equation}
	\left\{\begin{array}{lr}

		x^0 \in \mathbb{B}^n,  \\
		
		x^{t+1}_i = 	\left\{\begin{array}{lr}

		f_{i}(x^t) \text{ if } S^t = i, \in \mathbb{B}^n   ,\\
		
		x^{t+1}_i \text{ otherwise.} 
	\end{array}\right.
	\end{array}\right.
\end{equation}

Now, we show the link between asynchronous interations and Devaney's chaos. We introduce the function $F_f$ that is defined for any given application
$f: \mathbb{B}^n \to \mathbb{B}^n$ by $F_f : [[1;n]] \times \mathbb{B}^n \to \mathbb{B}^n$ such that:

\begin{equation}
	F_f(s,x)_j = \left\{\begin{array}{lr}

		f_{j}(x) \text{ if } j=s \in \mathbb{B}^n ,  \\
		
		x_j \text{ otherwise. }
	\end{array}\right.
\end{equation}

With such a notation, asynchronously obtained configurations are defined for times $t=0, 1, 2,...$ by:

\begin{equation}
	 \left\{\begin{array}{lr}

		x^0 \in \mathbb{B}^n \text{ and }  \\
		
		x^{t+1} = F_f(S^t, x^t) .
	\end{array}\right.
\end{equation}

Finally, iterations defined in Eq. 2 can be described by the following system,where $\sigma$ is the so-called shift function that removes the first term of the strategy (i.e., $S^0$).

\begin{equation}
	 \left\{\begin{array}{lr}

		x^0 = ((S^t)^{t\in \mathbb{N}} , x^0) \in [[1;n]]^{\mathbb{N}} \times \mathbb{B}^n,  \\
		x^{k+1} = G_f(X^k), \\
		\text{ where } \displaystyle \left( ((S^t)^{t\in \mathbb{N}},x) \right) = \displaystyle \left(\sigma((S^t)^{t\in \mathbb{N}},F_f(S^0, x)\right) . 
		
	\end{array}\right.
\end{equation}

The relation between $\Gamma(f)$ and $G_f$ is obvious: there exists a path from $x$ to $x^\prime$ in $\Gamma(f)$ if and only if there exists a strategy $s$ such that iterations of $G_f$ from the initial point $(s,x)$ reach the configuration $x^\prime$.

\begin{theorem}
Let $f: \mathbb{B}^n \to \mathbb{B}^n$. Iterations of $G_f$ are chaotic according to Devaney if and only if $\Gamma(f)$ is strongly connected.
\end{theorem}

\subsection{Chaotic iterations according to Devaney}

In this section, we stablish the link between such iterations and Devaney's chaos is finally presented at the end of this section. In what follows and for any function $f$,$f^n$ means that composition $f^o f^o ... ^of$ ($n$ times) and an \textbf{iteration} of a \textbf{dynamical system} is the step that consists in updating the global state $x^\prime$ at time $t$ with respect to a function $f$ s.t. $x^{t+1} = f(x^t)$. This definition consists of three conditions: topological transitivity, density of periodic points, and sensitive point dependence on initial conditions.

\paragraph{Topological transitivity} It's checked when, for any point, any neighbourhood of its future evolution eventually overlap with any other given region. This property implies that a dynamical system cannot be broken into simpler subsystems. Intuitively, its complexity does not allow any simplification.

\paragraph{Density of periodic points} As chaos needs some regularity to "counteracts" the effects of transitivity. In the Devaney's formulation, a dense set of periodic points is the element of regularity that a chaotic dynamical system has to exhibit. We recall that a point $x$ is a \textbf{periodic point} for $f$ of periodic $n \in \textbf{N}^{*}$ if $f^n (x) = x$. Then, the map $f$ is \textbf{regular} on the topological space $(\chi,\tau)$ if the set of periodic points for $f$ is dense in $\chi$ (for any $x \in \chi$, we can find at least one periodic point in any of its neighbours. Thus, due to these two properties, two points close to each other can behave in a completely different manner, leading to unpredictability for the whole system.

\paragraph{Sensitive dependence on initial conditions} Let us recall that $f$ has sensitive dependence on initial conditions if there exists $\delta > 0$ such that, for any $x \in \chi$ and any neighbourhood $V$ of $x$, there exist $y \in V$ and $n>0$ such that $(d(f^n(x),f^n(y))>\delta$. The value $\delta$ is called the \textbf{constant of sensitivity} of $f$.

\subsection{Chaotic Turing Machines}

Let $(w, i, q)$ be the current configuration of the Turing machine (Figure 2), where $w=\#^{-\omega}(0) . . . w(k)\#^{\omega}$ is the paper tape, $i$ is the position of the tape head, $q$ is used for the state of the machine, and $\tau$ is its transition function (the notations used here are well-known and widely used). We define $f$ by:

\begin{equation}
\begin{aligned}
&f(w(0) ... w(k),i,q) = (w(0) ... w(i - 1)aw(i + 1)w(k), i+1,q^\prime) \\
&\text{ if } \tau(q,w (i)) = (q^\prime, a,\rightarrow)
\end{aligned}
\end{equation}
\begin{equation}
\begin{aligned}
&f(w(0) ... w(k),i,q) = (w(0) ... w(i - 1)aw(i - 1)w(k), i+1,q^\prime) \\
&\text{ if } \tau(q,w (i)) = (q^\prime, a,\leftarrow)
\end{aligned}
\end{equation}

Thus the Turing machine can be written as an iterate function $x^{n+1} = f(x^n)$ on a well-defined set $\chi$, with $x^0$ as the initial configuration of the machine, which is $x^0=\{e_i = 1: e \in L_e\} \vee \{e_i = 0: e \notin L_e\}$ in our case. We denote by $\tau(S)$ the iterative process of the algorithm $S$. We show, experimentally, that our algorithm is capable of simulating a finite chaotic turing machine in a variety of hard instances of HCP with the emergence of chaotic iterations from its dynamics.

\section{SFCM-R+ algorithm} 

\subsection{Background}
In this section, we present the first  implementation of SFCM-R algorithm, called SFCM-R+, that relies on making the following conjecture hold, experimentally, for $H$ with a non-enforced parametrized polynomial-time runtime $log_\varepsilon V$.

\begin{conjecture}
Let \Gv. $H$ has a hamiltonian sequence when (1) $|x^{t}_k \cap  x^{t}_{m}| > 0$ for every $z^t_{k}$, (2) $z^t_{m}$ maps a  hamiltonian sequence, (3) $w^{*}=0$, (4) $\Gamma(f)$ is strongly connected.
\end{conjecture}

\begin{equation}
w^{*} = g(z^t, p) \text{ , } z^t = \displaystyle \left(\frac{|x^{t}_k \cap  x^{t}_{m}|}{|x^{t}_{m}|} \right)^{m}_{k=i}
\end{equation}
\\

This is done by treating minimal mapping constraints as being part of a stable manifold of $g(z^t,p)$, as the infinite number of unstable periodic orbits typically embedded in chaotic attractor could be taken advantage of for the purpose of achieving control by means of applying only very small perturbations in the system, which are related to enforcing ordering constraints progressively. By doing so, we can assume that there exists a nominal parameter value $p_0$, which is the desired orbit to be stabilized.

To every $x^t \in (x^{t}_{1},..., x^{t}_{n} )$ we associate an analog cost function $K(s) \in [0,1]$ via $K(s = x^t)= (max(\varepsilon(x^t)) - min(\varepsilon(x^t))^{-1} \sum (\varepsilon(x^t_{i}) - min(\varepsilon(x^t))$,$ 0 \leq \varepsilon(w) \leq |V|$. Defining the energy function $V(s,\varepsilon)$ associated with $F = x^t$ via $V(s,a) = \sum^{M}_{m=1} = K_m(s)$, we see that as long as $a_m > 0$,$\forall m \in \{1, ... , M\}$ we have $V(s^{*})=0$ if and only if $s^{*}$ is a solution to F . Thus, we are looking for the lowest/zero energy points of $V$ by assuming that asynchronous iterations starting from $L_e$ leads to the formation of a global attracotr of the fixed point attractor $x^{*}$ where $V(s^{*})=0$, which corresponds to the solution.   The search dynamics is as defined in is a gradient descent in $s$-space with a constrained ascent in the auxiliary $\varepsilon$-space.

\begin{equation}
\begin{array}{rcll} \displaystyle \frac{d s_i}{dt} = - \frac{\partial}{\partial s_i} V(\varepsilon,s),
 &&  i = 1, ... , N \end{array}
\end{equation}

We want to show that the dynamics is focused in the sense that the $x^t_{i}$ with the largest $K_m$ value violates the upper bound of energy of $V$, which is $\varepsilon_{MAX}$, and the dynamics drives exponentially fast to change such dominance until another $x^t_{i}$ happen to violates $\varepsilon_{MAX}$ and so on. Through experimentation, we show that the proposed system found at least one solution $s^{*}$ when one exists .

Here, $\Delta p$ cannot be changed directly as we can't have access to $x^{t}_{m}$ until the reconstruction process is done. Because of that, the algorithm uses a local search scheme in order to change $\Delta p$ indirectly by solving the following optimization by using the functions \textsc{Mapping} and \textsc{Reconstruct}, respectively (see ref for details). 

\begin{equation} 
\label{eq:3}
{\begin{array}{rcll} \text{minimize} \hphantom{00} && {\displaystyle F = \sum_{v \in V} \sum_{i=0}^x  v_{\displaystyle i_\varepsilon}} \\[12pt] \text{subject to} \hphantom{00}&& 0 \leq v_{\displaystyle i_\varepsilon} \leq |V| \end{array}}
\end{equation}

\begin{equation} 
{\begin{array}{rcll} \text{minimize} \hphantom{00} && {\displaystyle  \sum_{i=0}^{x} |H_i[V-v]_{\VH}| - |H_{i_ \VA} \cap H_i[V-v]_{\VH}|} \\[12pt] \text{subject to} \hphantom{00}&& {\displaystyle H_{i_\VA} \cap H_i[V-v]_{\VH} \neq \emptyset} \end{array}}
\end{equation}

\begin{equation} 
{\begin{array}{rcll} \text{maximize} \hphantom{00} && {\displaystyle \sum_{i=0}^{x} |P^{i}_{x_1}| + |P^{i}_{x_2}|} \\[12pt] \text{subject to} \hphantom{00}&& {\displaystyle |P^{i}_{x_1}| + |P^{i}_{x_2}| \leq |V|} \end{array}}
\end{equation}

The reconstruction process forms a continuous-time dynamical system. A continuous-time dynamical system as a \textit{deterministic algorithm} does have these features: 1) the search happens on an energy landscape $V = \sum \varepsilon z^t$, 2) it solves easy problems efficiently (polynomial time, both analog and discrete) and 3) it guarantees to find solutions to hard problems even for solvable cases where many other algorithms fail. According to experimental analysis, although it is not a mathematically proved polynomial cost algorithm, it seems to find solutions in continuous-time \textit{t} that scales polynomially on dataset, which is composed of a variety of hard HCP instances.

\usetikzlibrary{datavisualization}
\usetikzlibrary{datavisualization.formats.functions}

\subsection{Reconstruction phrase} 

In this section, the reconstruction phrase is explained.  The reconstruction task is done by the \textsc{Reconstruct}  function, that takes following parameters as input by reference:  \Gv,  $L_e$, $H^{*}$, $\phi$, $P_{x_1}$, $P_{x_2}$. The edge $\phi$ is a non-synchronized edge $(x_1 x_2) \in L_e$ where $x_1$ and $x_2$ are initially the last vertices of two expandable paths $P^\prime=(x_1)$ and $P^{\prime\prime}=(x_2)$, respectively. $P_{x_1}=P^\prime$ will be the current path we're expanding and $P_{x_2}=P^{\prime\prime}$, the other path. As for every $u$, $v$ must be added to either $P_{x_1}$ or $P_{x_2}$, $x_1$ and $x_2$ must be properly updated in order to represent the last vertices of $P_{x_1}$ and $P_{x_2}$, respectively. Every $H$ has the following property.

\begin{prop}{\emph{($|H^{n}|$ property)}}
The property $|H^{n}|$ indicates that $H^\prime$ is a component of $H^\prime \supset H[V-H_\VH]$ has $n$ vertices $w \in Z$ with $Z=\{ w^\prime \in V(H[V]) : (|N(w) \cap H_\VH|=1) \wedge (w^\prime \neq \VH)\}$. The value of $|H^{n}|$ is equal to $\alpha = \left|\bigcup_{w \in Z} \{N(w) \cap H_\VH\}\right|$; $|H_\VH^{n}|$ returns a set $\beta=\bigcup_{w \in Z} \{N(w) \cap H_\VH\}$. $|H^{n}|=0$ indicates that $H^\prime$ is a disconnected component of $H[V-H_\VH]$.
\end{prop}

The term \textit{expansion call} is used throughout this paper whenever we make a recursive call to \textsc{Reconstruct}. Every expansion call restores the initial state of both $H$ and $L_e$. Some conventions are used in this section.   The \textit{synchronized} edges will be written as $[v,u]$. The edge $[w,\square]$ is a synchronized edge $e \in L_e$ with $w \in e$. 
 
\begin{definition}
A synchronized edge is either: (1) a non-synchronized edge $(v,u)$ that got converted to $[v,u]$ by \textsc{Reconstruct}; or (2) an edge $[v,u]$ added to $L_e$ by \textsc{Reconstruct}.
\end{definition}

The notation $d^{*}(x)$ is used to represent the degree of a vertex $x$ of a scene $H^{*}$, which is a clone of  $H^\prime \supseteq H$ scene of the current state of \textsc{Reconstruct}, such that $V(H^{*})=V(H^\prime)$ and $E(H^{*})=L_e \cap E(H^\prime)$. 

$P_v(u)$ function is used by \textsc{Reconstruct} to pass through $H$ by using paths of $H^{*}$, starting from $v \in \{x_1,x_2\}$ until it reaches $z=u$ such that $d^{*}(z)=1$. During this process, it performs successive $H-v$ operation, converts edges from $(v,u)$ to $[v,u]$, and updates $P_v$. When $z$ is reached, it returns $z$.
$[v,u]$ cannot be removed from $H$ unless by undoing operations performed by reconstruction phrase. 

The goal of reconstruction phrase is to reconstruct a hamiltonian sequence (if it exists) by passing through $H$ in order to attach inconsistent $C_\VH$ components. If such hamiltonian sequence is reconstructed, $H^{*}$ will be a path graph corresponding to a valid hamiltonian sequence of the maximal $H \equiv G$. In order to do that, some edges may need to be added to $L_e$ to merge a component $H^{*}_{\prime}$ with $v \in V(H^{*}_{\prime})$ to another component $H^{*}_{\prime\prime}$ so that $P_v(u)$ can reach vertices $u \in V(H^{*}_{\prime\prime})$ properly.  If this path graph is found, it means that exist a sequence of $C_\VH$ components that convert $L_e$ to a valid hamiltonian sequence. Such sequence is part of the \textit{forbidden condition} of $A_0 = SFCM-R$.

\begin{definition}
Let \G be a graph. The Synchronization-based Forbidden Condition Mirroring (SFCM) algorithm is an algorithm with a configuration $g: W \times F \rightarrow A$, that consists of: (1) a finite set of scenes $W = W_0...W_n$,$W_0=G$, $W_0 \equiv ... \equiv W_n$, associated to each synchronizable forbidden condition in $F=F_0... F_n$; and (2) a pair $(w,f)$, with $w \in W$ and $f \in F$, associated to each mirrorable algorithm in $A = A_0 ... A_n$.
\end{definition}

\begin{definition}\emph{(Synchronizable forbidden condition)}
If $F_i \in F$ and $F_k \in F$ don't make the algorithms  $A_i \in A$ and $A_k \in A$, respectively, fail to produce a valid output, then both $F_i \in F$ and $F_k \in F$ are synchronizable forbidden conditions that will be synchronized eventually when both $A_i$ and $A_k$ are executed.
\end{definition}

In reconstruction phrase, a potential hamiltonian path $L_e$, which is the output of \textsc{Mapping}, is iteratively augmented by a local search scheme, one element at a time by \textsc{Select-First}. At each reconstruction iteration, $N(v)$ ordering can change depending on each $u^x \in N(v)$, $|\{ e \in L_e : (u \in N(v)) \wedge (u \in e)\}|$ and $d^{*}(u)$. $u^x$ is the priority of $u$, which is changed when $\varepsilon_{MAX}=\varepsilon_{u},V(s^{*})>0$. $N(v)$ ordering is changed by \textsc{Reorder} function, which returns $vN$. $vN$ is a set $S \supseteq N(v)$ In addition, $N(v)$ ordering can be changed by a shuffle operation before \textsc{R-Policy} reset the local priorities of reconstruction process in order to it not being stuck on local minimum. The current local search state is validated by \textsc{Valid-state}, that throws an error when one is found. This operations makes \textsc{Reconstruct} attempt to attach $\VH \in \nC$.

When a dead-end is found or every neighbour of $v$ is visited, we have $u = \textit{null}$. In this case,\textsc{Path-split} is called to change by reference $H$,$u$,$uN$,$vN$ and \textit{splitable}. This function is called when $P_{x_1}$ reaches dead-end and $P_{x_2}$ needs to be expanded through $vN$ to reach another valid dead-end, or throw an error exception otherwise. $P_{x_2}$ is expanded  by $P_{x_1 = x_2}$, with (1) $(x_1 x_2) \in L_e$ or (2) $x_2 \in V(H^\prime)$,$H^\prime \in S_1$. If both (1) and (2) hold, or $P_{x_2} = \emptyset$, an error is thrown. After this operation,\textit{splitable} is set to \textbf{true}. $vN$ is changed by \textsc{Rec-Node}, which takes $u$ and \textit{pass} as parameter.

\inicioAlgoritmo{Reconstructor + R-Policy}
\Input $L_e$, $\phi$, \textit{hc}, \textit{unrestrictedMode},  \textit{hcp}, \textit{longestPathFound}
\Output Set $L_e$ of synchronized edges
\Function{Reconstruct}{}

\State $S \gets (\emptyset)$
\State $\textit{pass} \gets \textit{true}$
\State $vN \gets \textsc{Rec-node}(x_1,  \textit{pass})$
\State $u \gets \textsc{Select-First}(vN)$

\While{reconstruction of $L_e$ is not done}
\If{$((u = \textit{null}) \wedge \textit{splitable})$} 
\State $u \gets \textsc{Path-Swap}(H, vN)$
\EndIf
\If{$u = \textit{null}$}
\State \textbf{throw} expand
\EndIf
\State $u_{LAST} \gets u$
\While{$d^{*}(u) \neq 1$}
\State $vN \gets \textsc{Rec-node}(u,  \textit{pass})$

\Try

	\If{\textsc{Valid-state}($H$, $\nC$)}
		\State $\nA \gets \nC \cup \nA $
		\State  $\nC \gets \emptyset$ 	
	\EndIf

\EndTry
\Catch{error} 
		\State $vN \gets \textsc{Reorder}(N(v))$
		\State \text{Undo $k$ states until: (1) $|S| = 1$ or }
		\State \text{(2)  $vN \sim w$, $w \in \nC$}
		\State $d^{*}(vN) \gets 1$.

\EndCatch
\State $u \gets \textsc{Select-First}(vN)$

\If{\textit{first}} $V \gets V \cup v$
\EndIf
\If {($u \neq \textit{null}$)}
             
     \State  $vN \gets \textsc{Rec-node}(u,  \neg\textit{pass})$
     \State $u \gets \textsc{Select-First}(vN)$
       
\Else

  \State \textsc{Path-split}($H$,$u$,$uN$,$vN$, \textit{splitable}) 	

\EndIf
	\EndWhile

    \If {$u_{LAST}^{i} = |V|$}
    \State \textbf{throw} expand 
    \EndIf

\If {$x_1 = \{\text{last visited } u \sim v\}$}
        
        \If{$P_{x_1} = \emptyset$} $vN \gets  \textsc{Rec-node}(x_1,  \textit{pass})$
          
          \If {($u \neq \textit{null}$)}
              \If{$d^{*}(u) = 1$}
     \State  $vN \gets \textsc{Rec-node}(u,  \neg\textit{pass})$
     \State $u \gets \textsc{Select-First}(vN)$
     \EndIf    
          \EndIf

      \EndIf
        \EndIf

  \State \textsc{Path-split}($H$,$u$,$uN$,$vN$, \textit{splitable})

\If {$|P_{x_1}| + |P_{x_1}| > \alpha$}
	\State $\textit{longestPathFound} \gets L_e$	
	\EndIf
	\EndWhile
\State \Return $L_e $
\EndFunction
\fimAlgoritmo{}

When \textit{pass} = \textbf{true}, $vN$ corresponds to $u \in (v,u)$, $u^{x \geq 1}$, $(v,u) \in L_e$. Otherwise, $u \in (w, u)$, $w\neq v$, $w^{x \geq 1}$.
The approach used by \textsc{R-Policy} to augment $L_e$ by local search is to find the $L_e$ of highest $a=|P_{x_1}| + |P_{x_2}|$ by using a well-structured dynamic in which global priorities and each $N(v),v \in V$ ordering is updated. In addition, the boolean variables \textit{hcp} and \textit{restrictedMode} are changed depending on specific criteria. When $\textit{restrictedMode} = \textbf{true}$, \textsc{R-Policy} makes \textsc{Reconstruct} attach only explicitly $C_\VH$ components. Otherwise,  \textsc{R-Policy} makes \textsc{Reconstruct} attach both isolated components $H^\prime \supset H$ with $|H^n|=0$ and explicitly $C_\VH$ components. When $\textit{hcp} = \textbf{true}$, \textsc{R-Policy} makes \textsc{Reconstruct} enforce $L_e$ to be a hamiltonian circuit. Otherwise, it makes \textsc{Reconstruct} enforce $L_e$ to be a hamiltonian path. When the local priorities of reconstruction process are reset, $L_e$ is set to \textit{longestPathFound} and the process starts again until it reaches the stop condition or a valid solution. In our case, the stop condition is (1) the maximum number of \textsc{R-Policy} runs with $L_e$ being from last \textsc{R-Policy} call parameter instead of \textsc{Mapping}, (2) $\textit{longestPathFound} = \textit{null}$ or (3) $\exists{v^{-1}}$. the priority $x$ of $w$ is represented by $w^x$. In this process, there may be some $w \in V$ with $d^{*}(w)=0$ that is included to $L_e$ in line 30 and 31. Notice that we use $\nA$, $\nJ$, $\nC$ in a slightly different manner from theoretical algorithm. $\nA$ is used in a way that prioritizes $H_\VH$ sets found when the reconstruction of a hamiltonian circuit is enforced by \textsc{Reconstruct} with \textit{restrictedMode} set to \textit{true} initially. $\nJ$ is not explicitly enforced in \textsc{Reconstruct}. In this implementation, $\nJ$ is  equivalent to $\nJ = \{w \in V: w^{x<1}\}$ since $w^{x>1}$ tend to be visited first. 

\begin{figure}[H]
\label{fig:2}
\centering
\includegraphics[scale=0.54]{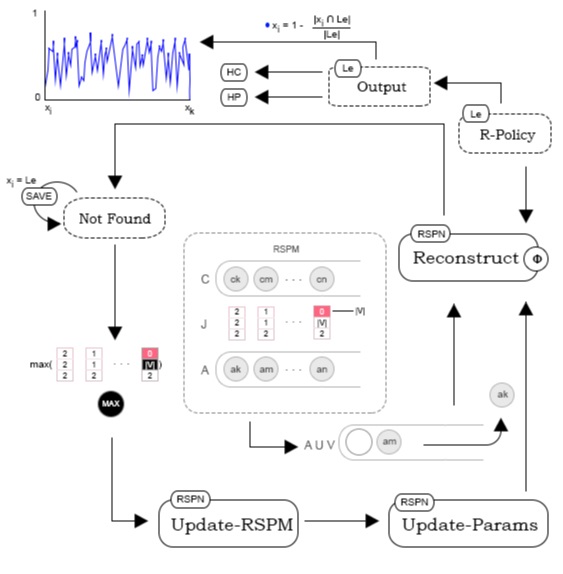}
\centering
\caption{Reconstruction phrase steps}
\end{figure}

\begin{figure}[H]
\label{fig:2}
\centering
\includegraphics[scale=0.4]{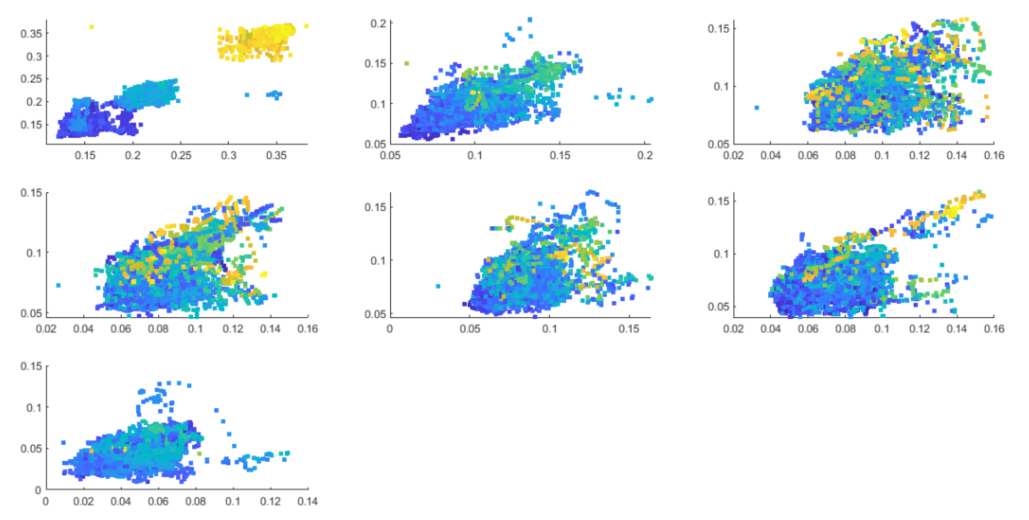}
\centering
\caption{Reconstructed phrase space of  signal $x_i=1-\frac{|x_i \cap L_e|}{|L_e|}$ from uniquely hamiltonian  Fleischner graph -  \textit{graph48}. $x=0$ is the solution. Notice how \textsc{Reconstruct} approaches $x=0$ through chaotic iterations (from left to right). Each image corresponds to a different \textsc{R-Policy} rounds, from a total of 7  rounds. See section 4.}
\end{figure}

\inicioAlgoritmo{R-Policy Controller}
\Input $L_e$,  \Gv, \textit{restrictedMode},  \textit{hcp}, \textit{longestPathFound}, \textit{round}
\Output Set $L_e$ of synchronized edges

\Function{R-Policy}{}

\State $\textit{lastLPF} \gets \emptyset$
\State $\textit{round} \gets \textit{round}+1$
\State $\textit{hc} \gets hcp$

\For {\textbf{each} $v \in V$}
\State Shuffle $N(v)$
\EndFor
\While{\textbf{true}}
\Try
\State$\phi \gets \text{random non-visited } e \in L_e$ 
\If {$\phi = \textit{null}$}
\State \textbf{throw} notfound
\EndIf
\State  $L_e \gets$ \textsc{Reconstruct($L_e$, $\phi$, \textit{hc}, \textit{restrictedMode},  \textit{hcp}, \textit{longestPathFound})}
\State \Return $L_e$

\EndTry
\Catch{expand}

\Try

\State $\phi \gets \text{first non-visited } e_v e_u \in Z $ ,$e_v \in Z$,$Z = \{\nA \cup V \}$.
\State $|v \in V|_\varepsilon \gets 0$ 
\If {$\phi = \textit{null}$}
\State \textbf{throw} notfound
\EndIf

\State  $L_e \gets$ \textsc{Reconstruct($L_e$, $\phi$, \textit{hc}, \textit{restrictedMode},  \textit{hcp}, \textit{longestPathFound})}
\State \Return $L_e$

\EndTry
\Catch{notfound}
\State $\textit{skip} \gets \textit{false}$
\For{\textbf{each}  $z$, $d^{*}(z)=0$}
 \State $d^{*}(z) \gets c \in \{1,2\}$
\EndFor
\State $\textit{freqMax} \gets max(|w|_\varepsilon ^{T})$
\State $|v \in V|_\varepsilon ^{T} = 0$, $Z \gets \emptyset$, $\nA,\nC \gets \emptyset$
\If{$\textit{freqMax}^{0} \wedge (d^{*}(freqMax)=2)$}
\State Remove the first element of $S=\{L_e \cap (freqMax,\square)\}$ from $L_e$
\EndIf

\If{$\textit{longestPathFound} = \textit{null}$}
\State \textbf{return} $L_e$
\EndIf

  \If {$(\textit{step2} \wedge \neg \exists{v^{-1}})$}

                 \If {($|\textit{lastLPF}| > |\textit{longestPathFound}|$)} 
                     \If {(\textit{restrictedMode})}
                       \State $\textit{freqMax}^n \gets \textit{freqMax}^{n-1} $ 
                       
                    \EndIf
                   \If{(\textit{hc})}

                      \State $\textit{restrictedMode}  \gets \textit{true}$

                   \Else

                      \State $\textit{restrictedMode}  \gets \textit{false}$

                   \EndIf
                
                   \State $\textit{skip} \gets \textit{true}$
                     \State $\textit{hc} \gets \textit{false}$
                 \EndIf
        
          \If{ $(\neg \textit{skip})$}
                   
                  \If {(\textit{restrictedMode})} 
                  \State $\textit{freqMax}^n \gets \textit{freqMax}^{n-1} $ 
                     \State $\textit{restrictedMode}  \gets \textit{false}$
                   \Else
                      \State $L_e \gets \textit{longestPathFound}$

                      \State $\textit{restrictedMode}  \gets \textit{true}$

                   \EndIf
                   
\EndIf
 
               \EndIf

                \If{$\exists{v^{-1}}$}

\State \textbf{return} $L_e$
\Else
\State $\textit{lastLPF} \gets \textit{longestPathFound}$
\EndIf

\EndCatch

\EndCatch

\Catch{notfound}
\State \textbf{return} $L_e$
\EndCatch

\EndWhile

\EndFunction

\fimAlgoritmo

In fact, when $w^{0}$, \textsc{Reconstruct} can't even attach $w \in C_\VH$ by calling $\textsc{Valid-State}$. $w^x$ is changed when $|w|^T_\varepsilon > |V|$, $\varepsilon_{MAX}=|V|$. $|w|^T_\varepsilon$ is the accumulated error counter of $w$ through i-th expansion calls. $|w|_\varepsilon$ is the error counter of $w$ before an expansion call. Because of that, it is set to zero before an expansion call by $|v \in V|_\varepsilon \gets 0$. When $|V|$ expansion calls are made, an  exception is thrown and $N(v)$ is shuffled. Notice that by not shuffling $N(v)$ before each expansion call and by choosing only the first $u \in N(v)$ locally, we're  relying the capacity of the algorithm of reorganize itself to transform $L_e$ into a hamiltonian path in a deterministic, well-structured manner. Notice that the shuffle operation only affects the order of attachments, not the dynamics itself. We show in the experimental section that even by using a shuffling operation, the behaviour of \textsc{Reconstruct} is chaotic in variety of hard HCP instances. Instead, such operation seems to be crucial to make the algorithm converge to solution.

\section{Experimental Analysis}
In our experiments, we ran the above algorithm for 94   sparse graphs of FHCP dataset (sparse graphs with at most 558 vertices) with a time limit of 18 hours on AMD EPYC 7501 32-Core Processor, 8 GB RAM, 4 vCPU. Solution is found for 91 out of 94 graphs due to hardware limitations. The difference between \textsc{Mapping} output and solution state of \textsc{Reconstruct} until convergence, is seen downsampled in Figure 3 and 4. The resulting convergence of \textsc{Reconstruct} is seen in Figure 4, which is the state that represents the difference between \textsc{Reconstruct} state and solution state. 

\begin{figure}[H]
\label{fig:2}
\centering
\includegraphics[scale=0.4]{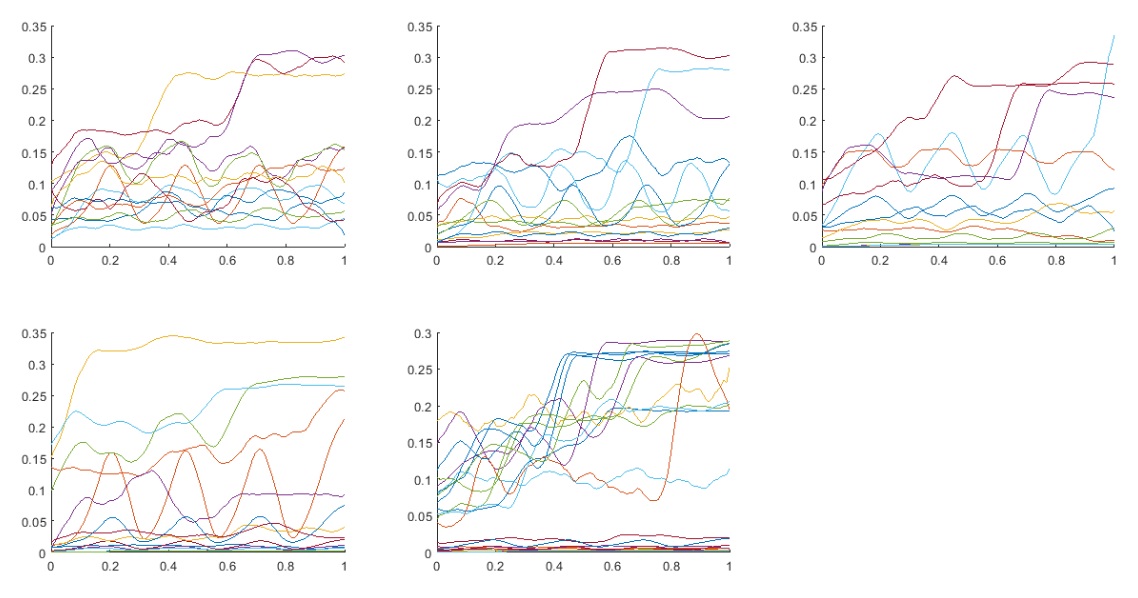}
\centering
\caption{\textsc{Mapping} process - $w_i=1-\frac{|w_i \cap L_e|}{|L_e|}$} 
\end{figure}

\begin{figure}[H]
\label{fig:2}
\centering
\includegraphics[scale=0.4]{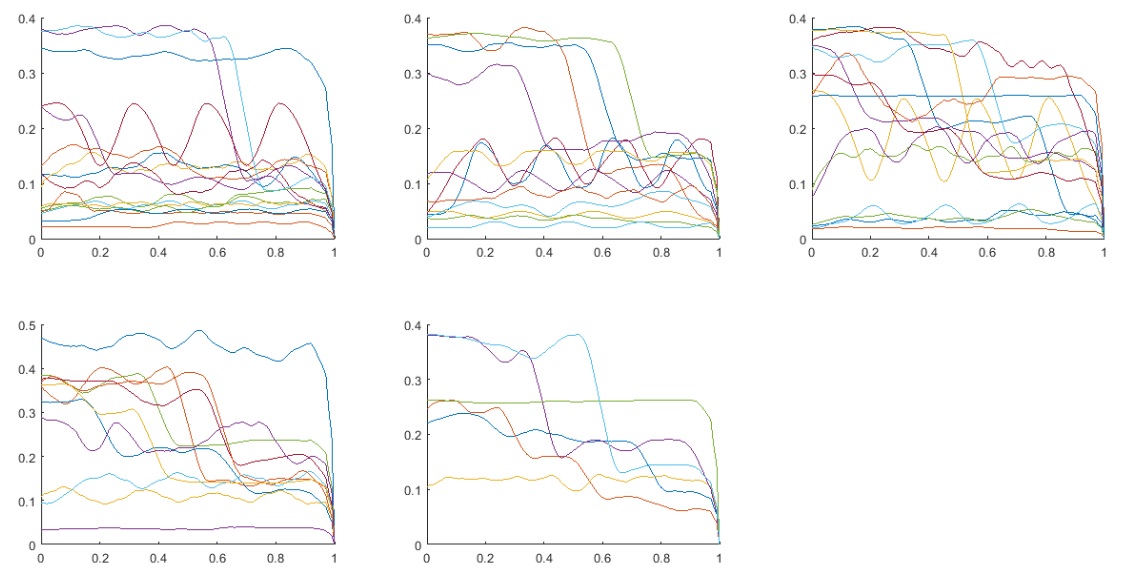}
\centering
\caption{\textsc{Reconstruct} process - $x_i=1-\frac{|x_i \cap L_e|}{|L_e|}$}
\end{figure}

The following figure shows Lyapunov and $0-1$ test for $x_i=1-\frac{|x_i \cap L_e|}{|L_e|}$, respectively, of \textsc{Reconstruct} states (starting from \textsc{Mapping} output until convergence) for 91 graphs used above, indicating the presence of chaotic behaviour for almost all of them.  These tests require a large number of datapoints to perform in a reliable manner. Because of that, these measurements are missing for a small fraction of instances, in which solution is found very fast, resulting in a very small $x_i=1-\frac{|x_i \cap L_e|}{|L_e|}$ .
The resulting performance of algorithm is seen in Figure 5. The worst case scenario for tested graphs is $log_{|R_\VH|} V \cong 5$ or $O(V^5 (V+E))$ for \textsc{Reconstruct}, and $log_{|M_\VH|} V \cong 2.5$ or $O(V^{2.5} (V+E))$ for \textsc{Mapping}  (See supplementary material).

\begin{figure}[H]
\label{fig:2}
\centering
\includegraphics[scale=0.4]{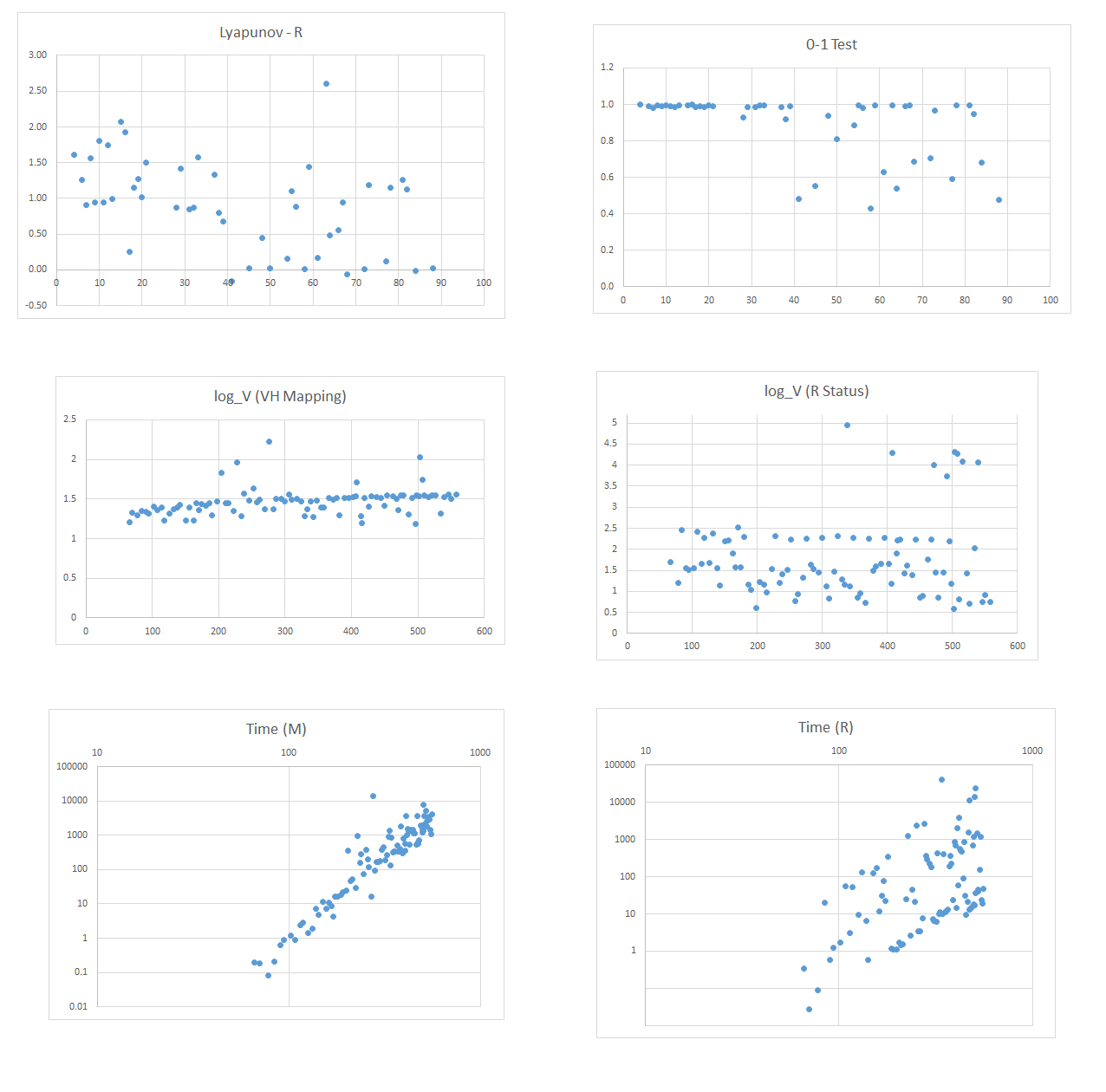}
\centering
\caption{Lyapunov, 0-1 test, $log_{R_{|\VH|}}$, $log_{M_{|\VH|}}$, running time (in seconds) of mapping phrase , running time (in seconds) of reconstruction phrase}.
\end{figure}

\section{Conclusion}

Our work shows emergent chaotic behaviour in hard instances of hamiltonian path problem. Elucidating the nature of NP-Complete problem is crucial to solve a variety of mathematical problems and can be promising for tackling real-world situations in which computational cost becomes a struggle.

\section{Supplementary material}
\subsection{Path splitter}

\inicioAlgoritmo{Path Splitter}
\Input $L_e$, $\phi$  \textit{hc}, \textit{restrictedMode},  \textit{hcp}, \textit{longestPathFound}
\Output \textbf{true}
\Function{Path-split}{}
\If {$\textit{splitable}$}
\If{$\textit{hc}$}
\State \textbf{throw} expand
\EndIf
\State $S \gets \{ H^\prime \supset H[V-H_\VH] : (|\{v  \cup N(v) \cup x_2 \cup N(x_2)\} \cap V(H^\prime)| = \emptyset)\wedge (0 \leq |H^n| \leq 1) \}$
\State $S_0 \gets \{ H^\prime \in S : (|H^n|=1) \wedge (|H_\VH^{n}|^{x<1}) \}$
\State $S_1 \gets \{ H^\prime \in S : |H^n|=0 \} $

 \If {$( (|S_0| > 1) \vee (|S_1| > 0))$} 
       \State \textbf{throw} expand
   
      \EndIf
      
      \If{($S_1 \neq \emptyset$)}
        \State $u^\prime \gets u^\prime \in V(H^\prime)$ with  $H^\prime \in S_1$
      \EndIf

\If{$( (P_{x_2} = \emptyset) \wedge \neg \textit{splitable})$}

 \State \textbf{throw} expand 
\EndIf

\If{$ H^\prime \in S_1$} 
\State $vN \gets \textsc{Rec-node}(u^\prime,  \textit{pass})$
\Else
\State $\textit{aux} = P_{x_1}$
\State $P_{x_1} = P_{x_2}$
\State $P_{x_2} = \textit{aux}$
\State $vN \gets \textsc{Rec-node}(x_2,  \textit{pass})$
\EndIf
\EndIf 
\State $u \gets \textsc{Select-First}(vN)$
\If{$uN = \textit{null}$} 
     \If{$d^{*}(u) = 1$}
     \State  $vN \gets \textsc{Rec-node}(u,  \neg\textit{pass})$
     \State $u \gets \textsc{Select-First}(vN)$
     \EndIf    
\Else 
\If{$ \textit{splitable}$}
\State $\textit{splitable} \gets \textit{false}$
 
 \Else
 \State \textbf{throw} expand 
\EndIf 
 \EndIf

\State \Return \textbf{true}
\EndFunction
\fimAlgoritmo

\subsection{Validate state}

\inicioAlgoritmo{Check if current state is valid}
\Input $v$, \Gv
\Output \textbf{true}
\State $\textit{iCounter} \gets 0$ 

\Function{Valid-state}{}
\State $X \gets H_\VH$
\If{$( (P_{x_2} = \emptyset) \wedge \neg \textit{splitable})$}

 \State \textbf{throw} expand 
\EndIf
\If{($\textit{hc}$)}
\State $S \gets (x_1 \in H_\VH) \cup (x_2 \in H_\VH)$

\If{$S \neq \emptyset$}
\State $\nC \gets \nC \cup S$
\State \textbf{throw} error
\EndIf

\EndIf

\State $V \gets V - v$

\State $S \gets \{ H^\prime \supset H[V-H_\VH] : (|\{v  \cup N(v) \cup x_2 \cup N(x_2)\} \cap V(H^\prime)| = \emptyset)\wedge (0 \leq |H^n| \leq 1) \}$
\State $S_0 \gets \{ H^\prime \in S : (|H^n|=1) \wedge (|H_\VH^{n}|^{x<1}) \}$
\State $S_1 \gets \{ H^\prime  \in S : (|H^n|=1) \wedge (|H_\VH^{n}|^{x \geq 1})\} $
\State $S_2 \gets \{ H^\prime \in S : |H^n|=0 \} $
\State $ S \gets S_0 \cup S_1 \cup S_2$
\State $\textit{valid} \gets \textbf{true}$ 

\For{\textbf{each} $H^\prime \in S$}
 \If{$|H^n|=1$}
 \State $\textit{ignore} \gets \textbf{false}$
 \If{$|H_\VH^{n}|^{x<1}$}
          \State $\textit{iCounter} \gets \textit{iCounter}+1$
         \EndIf
          \If{$((\textit{iCounter}>1) \wedge (|H_\VH^{n}|^{x<1})) $}
         
          \State $\textit{ignore} \gets \textbf{true}$
          \EndIf
           \If{$(((\neg \textit{ignored}) \wedge (|H_\VH^{n}|^{x<1})) \vee |H_\VH^{n}|^{x \geq 1})$}
           	
           	       \State $ |H_\VH^{n}|_\varepsilon \gets  |H_\VH^{n}|_\varepsilon + 1$
           	       \State $ |H_\VH^{n}|_\varepsilon ^{T} \gets  |H_\VH^{n}|_\varepsilon^{T} + 1$
           	         \State $V \gets V \cup v$
           	       \If {$|H_\VH^{n}|_\varepsilon > |V|$}
           	       \State $|H_\VH^{n}|_\varepsilon = 0$
           	       \State \textbf{throw} expand
           	       \EndIf
           \State $\nC \gets \nC \cup  |H_\VH^{n}| \cup V(H^\prime)$
    
           \State $\textit{valid} \gets \textbf{false}$ 
			        	
        	\EndIf
          \ElsIf{$|H^n|=0$}
      
           \State $\nC \gets \nC \cup V(H^\prime)$
            \State $\textit{valid} \gets \textbf{false}$ 
        	
          \EndIf
          \State $V \gets V \cup v$
          \If{$(\neg\textit{valid})$}
             \If{($\neg\textit{restrictedMode}$)}
         \State $\nC \gets X$
         
\EndIf
\State \textbf{throw} error
\EndIf

\EndFor

\State \Return \textit{valid}
\EndFunction
\fimAlgoritmo

\subsection{Create reconstruction node}

\inicioAlgoritmo{Create $vN$ node}
\Input $v$, \textit{pass}
\Output \textit{vN}

\Function{Rec-node}{}
\If {$v = -1$}
\State \textbf{throw} expand
\Else
\State $S_{L_e} \gets \{ e \in L_e : (u \in N(v)) \wedge (u \in e) \wedge (u^{x \geq 1})\}$
\State $S_{L_e} \gets \{ e \in S_{L_e}  : (v \in e) \wedge ( ((v \notin e) \wedge \neg\textit{pass} )) \vee ( (v \in e) \wedge \textit{pass} )))\}$ 
\State $S_1 \gets \{ e \in S_{L_e} : (d^{*}(u) = 1) \wedge  (u \in e) \}$
\State $S_2 \gets \{ e \in S_{L_e} : (d^{*}(u) = 2) \wedge  (u \in e) \}$
\State $S_{0} \gets \{ e \in E : (d^{*}(u) = 0) \wedge (u \notin S_{L_e}) \wedge (u \in e) \wedge (u^{x \geq 1})\}$
\State $vN \gets S_1 \cup S_2 \cup S_0$
\EndIf
\State \Return $vN$
\EndFunction
\fimAlgoritmo

\subsection{Reorder N(v)}

\inicioAlgoritmo{Reorder $N(v)$}
\Input $N(v)$ by reference
\Output $S$
\Function{Re-order}{}
\State $S \gets \{v \in N(v): v \text{ is non-visited}\}$
\State Order every $v \in S$ by $v_\varepsilon$ 
\State $S \gets \{v \in S : v^x = v^{x \geq 1}\} \cup \{v \in S : v^x = v^{x < 1}\} $
\State \Return $S$
\EndFunction
\fimAlgoritmo

\subsection{Path swap}

\inicioAlgoritmo{Path swap}
\Input $vN$,$H$,$P_{x_1}$,$P_{x_2}$ by reference
\Output $u$
\Function{Path-swap}{}
\State $P_{x_2} = P_{x_2} \cup \{ x_2 \}$
\State $vN \gets \textsc{Rec-node}(x_2,  \textit{pass})$
\State $\textit{aux} = P_{x_1}$
\State $P_{x_1} = P_{x_2}$
\State $P_{x_2} = \textit{aux}$
\State $u = \textsc{Select-First}(vN)$
\State $\textit{splitable} \gets \textit{false}$
\State \Return $u$
\EndFunction
\fimAlgoritmo

\subsection{Node selector}

\inicioAlgoritmo{Select first $u$}
\Input $vN$,$H$,$P_{x_1}$,$L_e$
\Output $u$
\Function{Select-first}{$vN$}
\State $V \gets V - v$
\State $u \gets \text{first } w \in vN$
\If{$(v,u) \notin L_e$}
\If{$d^{*}(u) > 1$}
\State $L_e \gets L_e - (\text{non-visited } e=\{w^\prime, u)\})$
\EndIf
\If{$d^{*}(v) > 1$}
\State $L_e \gets L_e - (\text{non-visited } e=\{w^{\prime\prime}, v\})$
\EndIf
\State $L_e \gets L_e \cup (v,u)$
\EndIf
\State $P_{x_1} \gets P_{x_1} \cup \{ u \}$ 
\State \Return $u$
\EndFunction
\fimAlgoritmo

\subsection{FHCP dataset}
In this section, the results of FHCP dataset is presented. $M_|VH|$ is the number of times articulation point is calculated in \textsc{Mapping} phrase. $R_|VH|$ is the number of times articulation point is calculated in \textsc{Reconstruct} phrase. $M_\varepsilon$ represents the number of errors of \textsc{Mapping} phrase. $R_\varepsilon$ represents the number of errors of \textsc{Reconstruct} phrase. $HC$ stands for hamiltonian cycle. $HP$ stands for hamiltonian path.

{\def\arraystretch{3}\tabcolsep=8pt
\begin{longtable}{|l|l|l|l|l|l|l|l|l|l|l|}
 \firsthline \textbf{Instance}   & 
  \multicolumn{10}{c |}{\textbf{SFCM-R+}} \\
	$H$ & $V$ & $E$ & $M_{|\VH|}$ & $R_{|\VH|}$ & $\mu_x$ & $M_{\varepsilon}$ & $R_{\varepsilon}$ & $log_{M_{|\VH|}} V$ & $log_{R_{|\VH|}} V$ & $L_{e}$ \\ \hline

graph1 & 66 & 99 & 0 & 264 & 95.161 & 156 & 1203 & 1,692877851 & 1,205315549 & HC \\ \hline
        graph2 & 70 & 106 & 0 & 0 & 100.00 & 285 & 0 & 1,204175754 & 1,330468521 & HC \\ \hline
        graph3 & 78 & 117 & 0 & 23 & 96.104 & 287 & 190 & 1,204355003 & 1,299026958 & HP \\ \hline
        graph4 & 84 & 127 & 0 & 12450 & 81.944 & 395 & 52486 & 2,452888991 & 1,349386814 & HP \\ \hline
        graph5 & 90 & 135 & 0 & 234 & 98.864 & 415 & 1079 & 1,552018969 & 1,339674111 & HP \\ \hline
        graph6 & 94 & 142 & 0 & 262 & 98.889 & 401 & 956 & 1,510524463 & 1,319298376 & HP \\ \hline
        graph7 & 102 & 153 & 44 & 293 & 92.929 & 668 & 1363 & 1,560537483 & 1,406340845 & HP \\ \hline
        graph8 & 108 & 163 & 0 & 19742 & 75.556 & 593 & 80749 & 2,413238824 & 1,363736745 & HP \\ \hline
        graph9 & 114 & 171 & 0 & 578 & 96.429 & 720 & 2469 & 1,64933302 & 1,389141795 & HP \\ \hline
        graph10 & 118 & 178 & 0 & 10550 & 91.071 & 344 & 52827 & 2,27950044 & 1,224277461 & HP \\ \hline
        graph11 & 126 & 189 & 0 & 724 & 99.187 & 586 & 3281 & 1,673993305 & 1,317813955 & HP \\ \hline
        graph12 & 132 & 199 & 0 & 25939 & 73.451 & 797 & 106999 & 2,37170685 & 1,368242002 & HP \\ \hline
        graph13 & 138 & 207 & 0 & 536 & 99.265 & 952 & 2195 & 1,56150623 & 1,391965073 & HP \\ \hline
        graph14 & 142 & 214 & 0 & 59 & 97.872 & 1184 & 274 & 1,132631959 & 1,427946079 & HP \\ \hline
        graph15 & 150 & 225 & 0 & 9605 & 95.27 & 483 & 60756 & 2,198248421 & 1,233379859 & HC \\ \hline
        graph16 & 156 & 235 & 14 & 23918 & 77.612 & 1168 & 74125 & 2,220559976 & 1,398663279 & HP \\ \hline
        graph17 & 162 & 243 & 3 & 2547 & 97.468 & 512 & 15962 & 1,902266913 & 1,226183096 & HC \\ \hline
        graph18 & 166 & 250 & 0 & 865 & 99.387 & 1607 & 3213 & 1,579612607 & 1,444080986 & HP \\ \hline
        graph19 & 170 & 390 & 54 & 178880 & 82.468 & 1098 & 431358 & 2,526324545 & 1,363224376 & HP \\ \hline
        graph20 & 174 & 261 & 0 & 769 & 99.419 & 1666 & 3317 & 1,57137607 & 1,437895194 & HP \\ \hline
        graph21 & 180 & 271 & 0 & 33460 & 70.779 & 1577 & 148764 & 2,29351346 & 1,417935831 & HP \\ \hline
        graph22 & 186 & 279 & 0 & 32 & 99.454 & 1884 & 408 & 1,150317371 & 1,443076545 & HP \\ \hline
        graph23 & 190 & 286 & 0 & 35 & 83.333 & 895 & 239 & 1,043727545 & 1,29536736 & HC \\ \hline
        graph24 & 198 & 297 & 0 & 1 & 99.492 & 2332 & 24 & 0,600963191 & 1,466355897 & HP \\ \hline
        graph25 & 204 & 307 & 696 & 33 & 98.995 & 17311 & 657 & 1,219920578 & 1,835065291 & HP \\ \hline
        graph26 & 210 & 315 & 0 & 31 & 99.517 & 2327 & 479 & 1,15421292 & 1,449818452 & HP \\ \hline
        graph27 & 214 & 322 & 0 & 18 & 93.659 & 2311 & 178 & 0,965674005 & 1,443434631 & HC \\ \hline
        graph28 & 222 & 333 & 0 & 462 & 96.804 & 1511 & 4080 & 1,538839298 & 1,354981326 & HP \\ \hline
        graph29 & 228 & 343 & 1880 & 50569 & 73.196 & 42885 & 274546 & 2,306516284 & 1,964560395 & HP \\ \hline
        graph30 & 234 & 351 & 10 & 40 & 96.552 & 1100 & 705 & 1,202165311 & 1,283712784 & HC \\ \hline
        graph31 & 238 & 358 & 46 & 514 & 90.129 & 5277 & 2304 & 1,414842665 & 1,5662809 & HC \\ \hline
        graph32 & 246 & 369 & 0 & 539 & 99.588 & 3396 & 4111 & 1,51152052 & 1,47681452 & HP \\ \hline
        graph33 & 252 & 379 & 110 & 31979 & 98.785 & 8181 & 225024 & 2,228794718 & 1,629385155 & HP \\ \hline
        graph34 & 258 & 387 & 0 & 1 & 99.609 & 3385 & 68 & 0,759866453 & 1,46356354 & HP \\ \hline
        graph35 & 262 & 394 & 0 & 2 & 99.231 & 3955 & 190 & 0,942295159 & 1,487468291 & HP \\ \hline
        graph36 & 270 & 405 & 0 & 216 & 98.885 & 2219 & 1700 & 1,32865718 & 1,376247089 & HP \\ \hline
        graph37 & 276 & 415 & 16483 & 65798 & 74.684 & 276396 & 316957 & 2,253668679 & 2,229305382 & HP \\ \hline
        graph38 & 282 & 423 & 0 & 1995 & 97.491 & 2286 & 10000 & 1,632487075 & 1,370912131 & HP \\ \hline
        graph39 & 286 & 430 & 0 & 1539 & 99.648 & 4778 & 5589 & 1,52556014 & 1,497841159 & HP \\ \hline
        graph40 & 294 & 441 & 0 & 960 & 99.659 & 5212 & 3703 & 1,445725919 & 1,505867655 & HP \\ \hline
        graph41 & 300 & 451 & 0 & 77867 & 75.697 & 4416 & 429150 & 2,273852806 & 1,471477853 & HP \\ \hline
        graph42 & 306 & 459 & 81 & 54 & 99.016 & 7442 & 610 & 1,120531772 & 1,557571828 & HP \\ \hline
        graph43 & 310 & 466 & 0 & 0 & 100.000 & 5172 & 112 & 0,822529313 & 1,490613958 & HC \\ \hline
        graph44 & 318 & 477 & 0 & 840 & 99.684 & 5867 & 4570 & 1,46254657 & 1,505904431 & HP \\ \hline
        graph45 & 324 & 487 & 0 & 109997 & 74.17 & 4974 & 664378 & 2,319183771 & 1,472471422 & HP \\ \hline
        graph46 & 330 & 495 & 0 & 103 & 99.696 & 1731 & 1647 & 1,277218898 & 1,285796762 & HP \\ \hline
        graph47 & 334 & 502 & 169 & 92 & 74.545 & 2984 & 836 & 1,157884247 & 1,376841479 & HP \\ \hline
        graph48 & 338 & 776 & 594 & 7,88E+11 & 97.612 & 5175 & 3,36E+12 & 4,953243132 & 1,468577575 & HP \\ \hline
        graph49 & 342 & 513 & 0 & 18 & 99.11 & 1703 & 700 & 1,122757983 & 1,275130765 & HC \\ \hline
        graph50 & 348 & 523 & 0 & 120365 & 75.51 & 5919 & 614744 & 2,277597408 & 1,484214332 & HP \\ \hline
        graph51 & 354 & 531 & 0 & 2 & 99.43 & 3553 & 148 & 0,851415825 & 1,392934744 & HP \\ \hline
        graph52 & 358 & 538 & 0 & 6 & 97.159 & 3575 & 271 & 0,952654944 & 1,391322942 & HC \\ \hline
        graph53 & 366 & 549 & 0 & 1 & 99.725 & 7790 & 70 & 0,719762689 & 1,518067553 & HP \\ \hline
        graph54 & 372 & 559 & 0 & 106058 & 74.516 & 6776 & 583811 & 2,243211807 & 1,490336278 & HP \\ \hline
        graph55 & 378 & 567 & 0 & 1371 & 99.733 & 7946 & 7170 & 1,495841483 & 1,513156535 & HP \\ \hline
        graph56 & 382 & 574 & 3 & 1466 & 75.532 & 2169 & 12995 & 1,593212755 & 1,292090504 & HP \\ \hline
        graph57 & 390 & 585 & 0 & 3997 & 99.742 & 8493 & 19536 & 1,656012604 & 1,516388713 & HP \\ \hline
        graph58 & 396 & 595 & 273 & 166848 & 77.848 & 8514 & 774169 & 2,26694641 & 1,512931027 & HP \\ \hline
        graph60 & 402 & 603 & 0 & 3484 & 99.749 & 9047 & 20936 & 1,659185348 & 1,519263117 & HP \\ \hline
        graph61 & 406 & 610 & 0 & 175 & 99.012 & 10028 & 1173 & 1,176640744 & 1,533898559 & HP \\ \hline
        graph62 & 408 & 936 & 3858 & 613735 & 76.8 & 29031 & 1,57E+11 & 4,288532001 & 1,709476422 & HP \\ \hline
        graph63 & 414 & 621 & 0 & 4709 & 74.817 & 2369 & 97387 & 1,906190423 & 1,289478266 & HP \\ \hline
        graph64 & 416 & 625 & 3 & 205491 & 51.338 & 1351 & 591838 & 2,20389353 & 1,195320271 & HP \\ \hline
        graph65 & 420 & 631 & 44 & 121867 & 72.571 & 9580 & 705717 & 2,229536705 & 1,517722896 & HP \\ \hline
        graph66 & 426 & 639 & 0 & 892 & 99.764 & 4987 & 5731 & 1,429305807 & 1,406338278 & HP \\ \hline
        graph67 & 430 & 646 & 0 & 873 & 96.262 & 10947 & 18619 & 1,621419213 & 1,533830834 & HC \\ \hline
        graph68 & 438 & 657 & 0 & 485 & 98.391 & 10860 & 4629 & 1,387667275 & 1,527870294 & HP \\ \hline
        graph69 & 444 & 667 & 9 & 135208 & 72 & 10059 & 835950 & 2,236994182 & 1,511891122 & HP \\ \hline
        graph70 & 450 & 675 & 0 & 2 & 99.553 & 5511 & 174 & 0,844466561 & 1,410075669 & HP \\ \hline
        graph71 & 454 & 682 & 3 & 0 & 100.00 & 12460 & 241 & 0,896487381 & 1,541374464 & HC \\ \hline
        graph73 & 462 & 693 & 0 & 5561 & 99.782 & 12230 & 49826 & 1,762884495 & 1,533949587 & HP \\ \hline
        graph75 & 468 & 703 & 0 & 162209 & 74.807 & 10113 & 943591 & 2,237540705 & 1,499816956 & HP \\ \hline
        graph76 & 471 & 1161 & 361 & 2,01E+11 & 70.759 & 4284 & 48100000000 & 3,996281903 & 1,358705969 & HP \\ \hline
        graph77 & 474 & 711 & 0 & 975 & 99.788 & 13473 & 7596 & 1,450263983 & 1,54327593 & HP \\ \hline
        graph78 & 478 & 718 & 0 & 0 & 100.00 & 13518 & 179 & 0,840796289 & 1,541714352 & HC \\ \hline
        graph80 & 486 & 729 & 0 & 1233 & 99.376 & 3143 & 8235 & 1,457459519 & 1,301755813 & HC \\ \hline
        graph81 & 492 & 739 & 7 & 172639 & 75.545 & 12258 & 11700000000 & 3,740087808 & 1,518749115 & HP \\ \hline
        graph82 & 496 & 745 & 0 & 212552 & 74.949 & 1527 & 787511 & 2,187459362 & 1,181176287 & HP \\ \hline
        graph83 & 498 & 747 & 0 & 224 & 99.798 & 14919 & 1453 & 1,172412581 & 1,547417436 & HP \\ \hline
        graph85 & 502 & 754 & 0 & 1 & 99.799 & 14465 & 37 & 0,580664111 & 1,540457182 & HP \\ \hline
        graph86 & 503 & 1241 & 28225 & 1,87E+11 & 81.002 & 296899 & 4,77E+11 & 4,322866737 & 2,02571572 & HP \\ \hline
        graph88 & 507 & 1251 & 2352 & 1,69E+11 & 77.459 & 51870 & 3,44E+11 & 4,264888908 & 1,743032301 & HP \\ \hline
        graph89 & 510 & 765 & 0 & 1 & 99.803 & 15614 & 148 & 0,801553265 & 1,54881089 & HP \\ \hline
        graph91 & 516 & 775 & 0 & 208087 & 74.941 & 13550 & 1,24E+11 & 4,089515015 & 1,523211527 & HP \\ \hline
        graph92 & 522 & 783 & 0 & 576 & 99.808 & 16174 & 7315 & 1,421884782 & 1,548685698 & HP \\ \hline
        graph93 & 526 & 790 & 0 & 2 & 99.618 & 16456 & 82 & 0,703353135 & 1,549557651 & HP \\ \hline
        graph94 & 534 & 801 & 0 & 14855 & 74.953 & 3827 & 338949 & 2,0275163 & 1,313585433 & HP \\ \hline
        graph95 & 540 & 811 & 0 & 196188 & 73.789 & 14544 & 1,27E+11 & 4,063764119 & 1,523456804 & HP \\ \hline
        graph97 & 546 & 819 & 0 & 1 & 99.816 & 17940 & 115 & 0,752850862 & 1,554082227 & HP \\ \hline
        graph99 & 550 & 826 & 0 & 5 & 98.903 & 12894 & 310 & 0,909135752 & 1,499942939 & HP \\ \hline
        graph100 & 558 & 837 & 0 & 9 & 98.022 & 18623 & 111 & 0,744665227 & 1,554648102 & HP \\ \hline
    \end{longtable}
  }


\begin{thebibliography}{}


\bibitem{1} Lima C (2019). SFCM-R: A novel algorithm for the hamiltonian sequence problem. \textit{arXiv preprint arXiv:1902.06713v4}

\bibitem{2} BAHI, Jacques M. et al. Neural networks and chaos: Construction, evaluation of chaotic networks, and prediction of chaos with multilayer feedforward networks. \textit{Chaos: An interdisciplinary Journal of Nonlinear Science}, v. 22, n.1, 2012.

\end{thebibliography}
\end{document}